\documentstyle[preprint,eqsecnum,aps]{revtex}
\tightenlines
\draft

\newcommand{\be}{\begin{equation}}
\newcommand{\ee}{\end{equation}}
\newcommand{\beal}{\begin{eqalign}}
\newcommand{\eeal}{\end{eqalign}}
\newcommand{\bea}{\begin{eqnarray}}
\newcommand{\eea}{\end{eqnarray}}
\newcommand{\bean}{\begin{eqnarray*}}
\newcommand{\eean}{\end{eqnarray*}}
\newcommand{\ba}{\begin{array}}
\newcommand{\ea}{\end{array}}

\newcommand{\ep}{\epsilon}
\newcommand{\Th}{\Theta}
\newcommand{\Ga}{\Gamma}
\newcommand{\ga}{\gamma}

\newcommand{\La}{\Lambda}

\newcommand{\de}{\delta}

\newcommand{\pa}{\partial}

\newcommand{\no}{\nonumber}

\newcommand{\res}{\mbox{res}}

\newcommand{\op}{{\mathcal{O}}}
\newcommand{\lan}{\langle}
\newcommand{\ran}{\rangle}

\begin{document}
\title
 { \sc Topological Field Theory approach to the \\
 Generalized Benney Hierarchy\/}
\author{
{\sc Jen-Hsu Chang$^1$ and Ming-Hsien Tu$^2$\/}\\
  {\it $^1$Department of Basic Course, Chung Cheng Institute of Technology,\\
National Defense University,\\
Tashi, Taoyuan, Taiwan\/}\\
   E-mail: changjen@math.sinica.edu.tw\\
       {\it $^2$Department of Physics, National Chung Cheng University,\\
   Minghsiung, Chiayi, Taiwan\/}\\
   E-mail:
 phymhtu@ccunix.ccu.edu.tw
 }
\date{\today}
\maketitle
\begin{abstract}
The integrability of the generalized Benney hierarchy with three  primary
 fields is investigated from the point  of view of two-dimensional topological
 field theories coupled to gravity. The associated primary free energy and
 correlation functions at genus zero are obtained via   Landau-Ginzburg
 formulation and the string equation is derived using the twistor construction
 for the Orlov operators. By adopting the approach of Dubrovin and Zhang  we
 obtain the genus-one corrections of the Poisson brackets of the generalized
   Benney hierarchy.
  \end{abstract}

\newpage
\section{Introduction}
Over the past ten years, the relationships between topological field theories (TFT)
 and integrable systems have attracted considerable interest \cite{D}.
It's Witten who realized that when the matter sectors of 2d TFT couple to gravity,
the systems can be described by integrable hierarchies. In particular, Witten \cite{W2} and
  Kontsevich \cite{K} showed that the partition function of the pure two-dimensional
 gravity, the simplest case, is equivalent to a particular tau function of the
 Korteweg-de Vries (KdV) hierarchy \cite{Di} characterized by the string equation.
 For nontrivial matter sectors such as $A_n$ topological minimal models \cite{Kr1}
 and $CP^1$ topological $\sigma$-model \cite{EY,EHY,EYY} the integrable hierarchies behind them
  correspond to the dispersionless Kadomtsev-Petviashvili (dKP) and dispersionless Toda (dToda)
  hierarchies, respectively. Furthermore, the genus-zero primary free energy of a 2d TFT  has been
 shown to satisfy the Witten-Dijkgraaf-Verlinde-Verlinde (WDVV) equation\cite{W2,DVV} of
   associativity which can be used to construct an integrable hierarchy of
    hydrodynamic type by taking into account the topological recursion relations
     of genus zero \cite{W2,DW}. In fact it has been realized that the WDVV equation is intimately
      related to TFT and integrable hierarchies.
       On one hand, 2d TFT can be classified by the solutions
      of WDVV equations in the sense that a particular solution of WDVV provides
      the primary free energy of some topological model. On the other hand, for example,
       one can construct a particular tau function of the dispersionless Lax hierarchy,
      which, in small phase space, turned out to be a solution of WDVV equation and satisfies,
       in full phase space, the dispersionless Virasoro constraints \cite{Kr1} .
   Therefore the WDVV equation, being a bridge, enables us to study dispersionless
  integrable hierarchies from the point of view of 2d TFT and vice versa.

         To formulate dispersionless Lax hierarchies (for a review, see \cite{TT1}),
          let $\La$ be  an algebra of Laurent series of the form
 \[
\La=\{A|A=\sum_{i=-\infty}^Na_ip^i\},
 \]
  with coefficients $a_i$
depending on an infinite set of variables $T_1\equiv X, T_2, T_3, \cdots$. Then
 one can define a Lie-bracket associated with $\La$ as follows \cite{LM}:
 \[
\{A, B \}=\frac{\pa A}{\pa p}\frac{\pa B}{\pa X}-\frac{\pa A}{\pa X}\frac{\pa B}{\pa p},
 \qquad A,B \in \La
 \]
which can be regarded as the Poisson bracket defined in the 2-dimensional phase space $(X, p)$.
 The algebra $\La$ with respect to the above Poisson bracket can be decomposed into
  the  Lie sub-algebras $\La=\La_{\ge k}\oplus \La_{< k}$ for $k=0,1,2$ where
$\La_{\ge k}=\{A\in \La| A=\sum_{i\ge k}a_ip^i\},~\La_{< k}=\{A\in\La| A=\sum_{i< k}a_ip^i\}$
 and we will use the notations : $\La_+=\La_{\ge 0}$ and $\La_-=\La_{< 0}$ for brevity.
  The Lax formulation of dispersionless hierarchies can be formally defined as
\[
\frac{\pa \La}{ \pa T_n}= \{(\La^{n/N})_{\ge k}, \La \},\quad n=1,2,3,\cdots
\]
where $\La^{1/N}$ is the $N$-th root of $\La$ defined by $(\La^{1/N})^N=\La$.
\begin{itemize}
\item For $k=0$, it's called dKP hierarchy \cite{KG} (see also \cite{TT}) and
its relations to TFT have been investigated in \cite{Kr1,AK,LP}.
\item For $k=1$, it's called dispersionless modified KP(dmKP)  hierarchy
\cite{Li,CT1,CT2} and its relations to TFT are under investigated \cite{CT3}.
 It is the purpose of this paper.
\item For $k=2$, it's called dispersionless Harry-Dym hierarchy
\cite{Li} and its relations to TFT are still open.
\end{itemize}
\indent In this paper we continue our previous work \cite{CT3} to study
the Lax operator of the form \cite{Li,CT2}
\begin{equation}
L=p^2+v^1p+v^2+v^3p^{-1},
 \label{laxop}
\end{equation}
which satisfies the non-standard Lax equations
\begin{equation}
\frac{\pa L}{\pa T_n}=\{L^{n/2}_{\geq 1}, L \}.
 \label{laxeq}
\end{equation}
The first few nontrivial equations are
\begin{eqnarray}
\frac{\pa}{\pa T_2}\left( \ba{c} v^1
\\ v^2\\ v^3 \ea \right) &=& \left( \ba{c} 2v^2_X\\
v^1v^2_X+2v^3_X\\ (v^1v^3)_X \ea \right),\no\\ 8\frac{\pa}{\pa T_3}
 \left( \ba{c} v^1 \\ v^2\\ v^3 \ea \right) &=&
\left( \ba{c} -3(v^1)^2v^1_X+12(v^2v^1)_X+24v^3_X\\
 12v^1_Xv^3+24v^1v^3_X+12v^2v^2_X+3(v^1)^2v^2_X\\
12(v^2v^3)_X+3((v^1)^2v^3)_X \ea \right),\no
\end{eqnarray}
etc. They are multi-variable generalization of the Benney equation \cite{B} which
describes long wave in nonlinear phenomena and has very rich integrable structures
\cite{LM,KM,Zak,EYR,FS,GiT}. Thus the set of the above equations is referred to
as the generalized Benney hierarchy.

Recently, the bi-Hamiltonian structure associated with the generalized Benney hierarchy
has been obtained from those structure of dmKP via truncation \cite{Li,CT2}.
 Starting from the bi-Hamiltonian structure the primary free energy of the corresponding
  2d TFT can be constructed. The key idea relies on the fact that the Poisson brackets
   both are of hydrodynamic type introduced by Dubrovin and Novikov (DN) \cite{DN1} and
    form a flat pencil (see below),
    in which, the geometric setting is provided by Frobenius manifolds\cite{Du4,Du2,Du3}.
   One way to define such manifolds is to construct a function $F(t^1, t^2, \cdots, t^m)$
    such that the associated functions,
\begin{equation}
c_{\alpha\beta\gamma}= \frac{\pa^3 F(t)}{ \pa t^\alpha \pa t^\beta \pa t^\gamma},
 \label{df}
 \end{equation}
 satisfy the following conditions \cite{Du2}:
\begin{itemize}
\item
 The matrix $\eta_{\alpha\beta}=c_{1\alpha\beta}$ is constant and
non-degenerate  (for the discussion of degenerate cases, see \cite{S}).
\item The functions
$c_{\beta\gamma}^{\alpha}=\eta^{\alpha\epsilon}c_{\epsilon\beta\gamma}$
 define an associative and commutative algebra with a unity element. The
associativity will give a system of non-linear partial differential equations for $F(t)$
\begin{equation}
\frac{\pa^3 F(t)}{\pa t^{\alpha} \pa t^{\beta} \pa t^{\lambda}} \eta^{\lambda \mu}
 \frac{\pa^3 F(t)}{\pa t^{\mu} \pa t^{\ga}
\pa t^{\sigma}} = \frac{\pa^3 F(t)}{\pa t^{\alpha} \pa t^{\ga} \pa t^{\lambda}}
 \eta^{\lambda \mu} \frac{\pa^3 F(t)}{\pa
t^{\mu} \pa t^{\beta} \pa t^{\sigma}}
 \label{WDVV}
\end{equation}
which is the so-called WDVV equations \cite{W2,DVV} arising from TFT (see sec. 3).
\item The functions $F$ satisfies a quasi-homogeneity condition, namely,
 $ {\mathcal L\/}_{E} F= d_{F}F+ (\mbox{quadratic~ terms}),$
where $E$ is known as  the Euler vector field.
\end{itemize}
 On the other hand, given any solution(or primary free energy) of the WDVV equation,
  one can construct a Frobenius manifold ${\mathcal M\/}$ associated with it.
   On such a manifold one may interpret $\eta^{\alpha\beta}$ as a flat metric and
    $t^{\alpha}$ the flat coordinates.
   The associativity can be used to defines a Frobenius algebra on each tangent space
    $T^t {\mathcal M\/}$. Denoting the multiplication of the algebra as $u \cdot v$
   then one may introduce a second flat metric on ${\mathcal M\/}$
defined by
 \begin{equation}
  g^{\alpha\beta}=E(dt^\alpha\cdot  dt^\beta),
   \label{sec}
 \end{equation}
 where $dt^\alpha\cdot  dt^\beta=c_{\gamma}^{\alpha\beta}dt^{\gamma}=
 \eta^{\alpha \sigma}c_{\sigma\gamma }^{\beta}
 dt^{\gamma}$. This metric, together with the original metric $\eta^{\alpha\beta}$,
  define a flat pencil (i.e, $\eta^{\alpha\beta}+\lambda g^{\alpha\beta}$ is flat
   as well for any value of $\lambda$) which corresponds to the compatible condition
   in integrable systems and thus  enables one to obtain a bi-Hamiltonian
    structure from a Frobenius  manifold ${\mathcal M\/}$.
 The corresponding Hamiltonian densities are defined recursively by the
formula \cite{Du2}
 \begin{equation}
  \frac{\pa^2 h_{\alpha}^{(n+2)}}{\pa t^\beta \pa
t^\gamma}=c_{\beta\gamma}^\sigma \frac{\pa h_{\alpha}^{(n)}}{\pa t^\sigma},\qquad
h_{\alpha}^{(0)}=\eta_{\alpha \beta}t^{\beta}
 \label{rec}
 \end{equation}
  where the subscript $\alpha(=1,2, \cdots, m)$ labels the primary fields and the
  superscript $n \geq 0$, the level of gravitational descendants.

  Our paper is organized as follows: In next section, we shall compute
  $c_{\alpha\beta}^\gamma$ from the generalized Benney hierarchy by using (\ref{rec}).
  We shall show that an additional hierarchy generated by Hamiltonians with logarithmic
   type emerges, which together with the ordinary Benney hierarchy are identified as
    renormalization group flows
    in genus-zero TFT coupled to 2d topological gravity. Then, in section 3, based on
  the fact that the dispersionless Lax operator (\ref{laxop}) can be viewed as a
  superpotential in  Landau-Ginzburg (LG) formulation of TFT \cite{V,DVV},
 we define  the fundamental correlation functions and derive their dynamic flows
 from the genus-zero topological recursion relations \cite{W2,DW} of the
associated TFT. In section 4, to establish the string equation describing the gravitational
 effect, we construct the twistor data for the generalized Benney hierarchy by using
  the Orlov operator of the dmKP hierarchy\cite{CT1}, as we have done in \cite{CT3}.
    In section 5, we shall show that the Poisson brackets of the generalized Benney
    hierarchy can be unambiguously quantized by using the Dubrovin and Zhang (DZ)
approach to bi-Hamiltonian structure in 2d TFT \cite{DZ}. To do so, we compute
 the associated $G$-function, which is independent of $t_1$ and satisfies
  the following Getzler equations \cite{G}
\begin{eqnarray}
&&\sum_{1 \leq \alpha_1,\alpha_2, \alpha_3, \alpha_4 \leq  3} z_{\alpha_1}
z_{\alpha_2}z_{\alpha_3}z_{\alpha_4} \left(3c_{\alpha_1 \alpha_2}^{\mu}
c_{\alpha_3 \alpha_4}^{\nu} \frac{\pa^2 G}{\pa t^{\mu} \pa
t^{\nu}}-4c_{\alpha_1 \alpha_2}^{\mu}
c_{\alpha_3 \mu}^{\nu} \frac{\pa^2 G}{\pa t^{\alpha_4} \pa t^{\nu}}
-c_{\alpha_1 \alpha_2}^{\mu}c_{\alpha_3 \alpha_4 \mu}^{\nu} \frac{\pa G}{ \pa t^{\nu}}\right.\no\\
&&\left. +2c_{\alpha_1 \alpha_2 \alpha_3}^{\mu}c_{ \alpha_4 \mu}^{\nu} \frac{\pa G}{ \pa t^{\nu}}
+\frac{1}{6}c_{\alpha_1 \alpha_2 \alpha_3}^{\mu}c_{\alpha_4 \mu \nu}^{\nu}
+\frac{1}{24}c_{\alpha_1 \alpha_2 \alpha_3 \alpha_4}^{\mu}
c_{ \mu \nu}^{\nu}- \frac{1}{4}c_{\alpha_1 \alpha_2 \nu}^{\mu}
c_{ \alpha_3 \alpha_4 \mu}^{\nu} \right)=0,
\label{geq}
\end{eqnarray}
with $c_{\alpha_1 \alpha_2 \alpha_3 \alpha_4}^{\mu}=\pa^2 c_{\alpha_1
\alpha_2 }^{\mu}/\pa t^{\alpha_3} \pa t^{\alpha_4}$, to
evaluate the genus-one free energy ${\mathcal F\/}_1$.
 Section 6 is devoted to the concluding remarks.

 \section{Integrability and Primary Free energy}
In this section, we shall identify the primary free energy associated with
 the generalized Benney hierarchy from its integrability and construct the
  additional flows generated by the logarithmic Hamiltonians. According to
  the theory of Frobinus manifold, our starting point is the bi-Hamiltonian
  structure of the generalized Benney hierarchy, which can be constructed
  from those structure of the dmKP hierarchy via the Dirac reduction and
  thus the Lax flows (\ref{laxeq}) have the following bi-Hamiltonian
  description \cite{Li,CT2}:
 \[
 \frac{\pa {\mathbf{v}}}{\pa T_n}=J_1\cdot\frac{\de H_{n+2}}{\de{\mathbf{v}}}
 =J_2\cdot\frac{\de H_n}{\de{\mathbf{v}}}
 \]
  where ${\mathbf{v}}^T=(v^1,v^2,v^3)$,
  $(\de H_i/\de {\mathbf{v}})^T=(\de H_i/\de v^1, \de H_i/\de v^2, \de
H_i/\de v^3)$ and
   \be
    J_1= \left( \ba{ccc} 0 & 0 &
2\pa \\ 0 & 2\pa & v^1\pa\\ 2\pa & \pa v^1 & 0 \ea \right),\qquad
 J_2= \left( \ba{ccc} 6\pa & 4\pa v^1 & 2\pa v^2 \\ 4v^1\pa
& v^2\pa+\pa v^2+2v^1\pa v^1 & 2\pa v^3+v^3\pa+v^1\pa v^2 \\
2v^2\pa & \pa v^3+2v^3\pa+v^2\pa v^1 & \pa
v^1v^3+v^1v^3\pa \ea \right)
 \ee
with Hamiltonians defined by
\[
H_n=\frac{2}{n}\int \res( L^{n/2})
\]
where $\pa= \pa/\pa X$ and $\res(\Lambda$) is the coefficient of
$p^{-1}$ of $\Lambda$. We list some of them:
\bean
 H_1&=&\int\left[v^2-\frac{1}{4}(v^1)^2\right],\\
  H_2&=&\int v^3,\\
H_3&=&\int \left[ \frac{1}{2}v^1v^3+\frac{1}{4}(v^2)^2-
\frac{1}{8}v^2(v^1)^2+\frac{1}{64}(v^1)^4\right], \no\\ H_4&=&\int v^2v^3,\\
 H_5&=&\int\left[-\frac{1}{512}(v^1)^6+\frac{3}{128}(v^1)^4v^2-\frac{1}{16}(v^1)^3v^3
-\frac{3}{32}(v^1)^2(v^2)^2+\frac{3}{4}v^1v^2v^3+\frac{1}{8}(v^2)^3+\frac{3}{4}(v^3)^2\right],\\
H_6&=&\int [v^1(v^3)^2+v^3(v^2)^2].
\eean
To find the primary free energy associated with the generalized Benney hierarchy, we have to
investigate the geometrical meaning of the Poisson brackets $J_1$ and
$J_2$ which can be written as the DN type \cite{DN1}
\[
J_1^{ij}=\eta^{ij}\pa+\ga^{ij}_{k}v^k_x,\qquad
J_2^{ij}=g^{ij}\pa+\Ga^{ij}_{k}v^k_x
\]
with $J_1^{ij}=\pa J_2^{ij}/\pa v^2$ where $\ga_k^{ij}$ and $\Ga_k^{ij}$ are the contravariant
Levi-Civita connections of the contravariant flat metrics $\eta^{ij}$ and $g^{ij}(v)$
respectively.  Let's introduce the flat coordinates
\begin{equation}
t^1=v^2-\frac{1}{4}(v^1)^2, \qquad t^2=v^3, \qquad t^3=v^1.
 \label{flat}
\end{equation}
so that  $\eta^{ij}$ and $g^{ij}$ can be expressed as follows:
\begin{equation}
\eta^{ij}(t)= \left(
 \ba{ccc}
 2& 0 & 0 \\
 0& 0 & 2\\
 0&2&0
   \ea
    \right),
    \qquad
    g^{ij}(t)=\
     \left(
 \ba{ccc}
  2t^1 & 3t^2& t^3 \\
  3t^2 & 2t^2t^3 & 2t^1+\frac{1}{2}(t^3)^2\\
  t^3 & 2t^1+\frac{1}{2}(t^3)^2 & 6
    \ea
    \right),
     \label{metric}
\end{equation}
with $\pa g^{ij}/\pa t^1= \eta^{ij}.$ It is easy to show that
$t^{\alpha}$ are just  the densities of the Casimirs invariants for the
first structure $J_1$.

Now we are in a position to compute the structure coefficients
$c_{\beta\gamma}^{\sigma}$ of the generalized Benney hierarchy. Choosing
$h_1^{(0)}=\eta_{1 \beta}t^{\beta}= t^1/2$ and
$h_3^{(0)}=\eta_{3 \beta}t^{\beta}= t^2/2$, the equations
(\ref{rec}) implies
\begin{eqnarray}
 h_{2n-1}&=&\frac{2^{n-1}}{(2n-1)!!}\res L^{n-1/2}, \qquad
h_1=h_1^{(0)}=\frac{t^1}{2},
   \label{for1}\\
 h_{2n}&=&\frac{2^{n-1}}{(2n)!!}\res L^n, \qquad
h_2=h_3^{(0)}=\frac{t^2}{2}.
 \label{for2}
\end{eqnarray}
with non-zero coefficients
\bean
c^1_{11}&=&1,\quad c^1_{23}=1,\\
c^2_{12}&=&1, \quad c^2_{23}=\frac{t^3}{2}, \quad c^2_{33}=\frac{t^2}{2},\\
c^3_{13}&=&1,\quad c^3_{22}=\frac{2}{t^2},\quad c^3_{33}=\frac{t^3}{2}.
\eean
Then from (\ref{df}), we immediately obtain the free energy
\[
F(t^1,t^2,t^3)=\frac{1}{12}(t^1)^3+\frac{1}{2}t^1t^2t^3+\frac{1}{24}(t^3)^3t^2
+\frac{1}{2}(t^2)^2\left(\log t^2-\frac{3}{2}\right)
\]
which is indeed a finite-dimensional solution to the WDVV equations (\ref{WDVV}).
Also, from (\ref{sec}) and  (\ref{metric}), we have the associated Euler
vector field
\[ E=t^1\frac{\pa }{\pa t^1}+\frac{3}{2}t^2\frac{\pa}{\pa
t^2}+\frac{1}{2}t^3\frac{\pa}{\pa t^3},\]
which implies the quasi-homogeneity condition: $
{\mathcal{L}}_E F(t)=3F(t)+\frac{3}{4}(t^2)^2$.

 Next we turn to the hierarchy corresponding to $\alpha=2$ in (\ref{rec}) with
$h_2^{(0)}=\eta_{2\beta}t^{\beta}=t^3/2$ and obtain
\begin{eqnarray*}
 \bar{h}_0&=& \frac{t^3}{2},\\
  \bar{h}_{1}&=& \frac{(t^3)^3}{24}+\frac{1}{2}t^1t^3+t^2\left(\log
t^2-\frac{5}{2}\right),\\
 \bar{h}_{2}&=&
\frac{(t^3)^5}{320}+\frac{t^1(t^3)^3}{24}+\frac{(t^1)^2t^3}{4}
 +\frac{1}{4}(t^3)^2t^2\left(\log t^2-\frac{3}{2}\right)+t^1t^2\left(\log t^2-\frac{5}{2}\right),
\end{eqnarray*}
As in the work \cite{CT3}, the Hamiltonian densities $h_2^{(n)}=\bar {h}_n$
can be expressed as
\begin{equation}
\bar{h}_{n}=\frac{3}{2n!}\res [L^n(\log L-c_n)],
\label{for3}
\end{equation}
where we use the following prescription for $\log L$
\bea
  \log L&=&\log(p^2+v^1p+v^2+v^3p^{-1}) \no\\
 &=&\frac{1}{3}\log [p^2(1+v^1p^{-1}+v^2p^{-2}+v^3p^{-3})]+
 \frac{2}{3}\log \left[v^3p^{-1}\left(1+\frac{v^1}{v^3}p+\frac{v^2}{v^3}p^{2}
 +\frac{1}{v^3}p^{3}\right)\right]\no\\
 &=&\frac{2}{3}\log v^3+\frac{1}{3}\log
(1+v^1p^{-1}+v^2p^{-2}+v^3p^{-3})+\frac{2}{3}\log
 \left(1+\frac{v^1}{v^3}p+\frac{v^2}{v^3}p^{2}+\frac{1}{v^3}p^{3}\right),
 \label{logl}
\eea
and $c_n-c_{n-1}=1/n,~c_0=1.$ The Lax and bi-Hamiltonian formulations
corresponding to $\bar{T}_n$-flows can be summarized as
\bea
\frac{\pa L}{\pa \bar{T}_n}&=&2\{\bar{B}_n, L\}, \qquad
\bar{B}_n=[L^n(\log L-c_n)]_{\geq 1}\no\\
&=&\{\bar{H}_{n+1}, L\}_1=\{\bar{H}_n,L\}_2,\qquad
\bar{H}_n=2(n-1)!\int \bar{h}_n
\label{logflow}
\eea
Notice that the additional flows  generated by the Hamiltonians $\bar{H}_n$ are compatible with
those flows of the ordinary Benney hierarchy (\ref{laxeq}). We will see
later that the forms (\ref{for1}) , (\ref{for2}) and (\ref{for3}) can be
used to formulate the generalized Benney hierarchy as a 2d TFT coupled to gravity
via LG description.

Before ending this section we would like to remark that the second Poisson bracket $J_2$
in fact reveals the classical realization of $w_3$-$U(1)$-Kac-Moody algebra \cite{CT2}
with $t^1$ being the Diff $S^1$ tensor of weight 2, and $t^2$ and $t^3$, tensors of weights
3 and 1, respectively. This is due to the fact that the Diff $S^1$ flows are just the Hamiltonian
flows generated by the Hamiltonian $H_1=\int t^1/2$.

\section{Landau-Ginzburg formulation}

 In this section we shall set up the correspondence between the Benney hierarchy
  and its associated TFT at genus zero. Let us first briefly recall some
   basic notions in TFT for convenience.

 A topological matter theory can be characterized by a set of BRST invariant observables
 $\{\op_1, \op_2,...\}$ with couplings $\{T^\alpha\}$ where $\op_1$ denotes the
  identity operator. If the number of observables is finite the theory is called
  topological  minimal model and the observables are referred to the primary fields.
  When the theory couples to gravity, a set of new observables emerge as gravitational
 descendants $\{\sigma_n(\op_\alpha), n=1,2,\cdots\}$ with new coupling constants
  $\{T^{\alpha,n}\}$.  The identity operator $\op_1$ now becomes the puncture operator $P$.
   For convenience we shall identify the primary fields  $\op_\alpha$ and the
   coupling constants $T^\alpha$ to  $\sigma_0(\op_\alpha)$ and $T^{\alpha,0}$, respectively.
 As usual, we shall call the space spanned by $\{T^{\alpha,n}, n=0,1,2,\cdots\}$ the full
 phase space  and  the subspace parametrized by $\{T^\alpha\}$ the small phase space. Hence,
 for a topological minimal model, the small phase space is finite-dimensional.

 For a topological model the most important quantities are correlation functions which
 do not depend on coordinates and describe the underlying topological properties of the manifold.
  The generating function of correlation functions is the full free energy defined by
 \[
{\mathcal F\/}(T)=\sum_{g=0}^\infty {\mathcal F\/}_g(T)=\sum_{g=0}^\infty
 \lan e^{\sum_{\alpha,n} T^{\alpha,n}\sigma_n(\op_{\alpha})}\ran_g
 \]
 where $\lan\cdots\ran_g$ denotes the expectation value on a Riemann surface of
  genus $g$ with respect  to a classical action.
   In the subsequent sections, we will omit the exponential factor
 without causing any confusion. Therefore a generic $m$-point correlation function
  can be calculated as follows
\[
 \lan\sigma_{n_1}(\op_{\alpha_1})\sigma_{n_2}(\op_{\alpha_2})\cdots\sigma_{n_m}
 (\op_{{\alpha}_m})\ran_g=\frac{\pa^m {\mathcal F\/}_g}{\pa T^{\alpha_1,n_1}
 \pa T^{\alpha_2,n_2}\cdots\pa T^{\alpha_m,n_m}}.
 \]
 In the following, we shall restrict ourselves to the trivial topology, i.e.,
  the genus-zero sector ($g=0$) since this part is more relevant to dispersionless
   integrable hierarchies. In particular, the genus-zero free energy restricting
   on the small phase space is the primary free energy defined by
   \[
 {\mathcal F\/}_0|_{T^\alpha=t^\alpha, T^{\alpha,n\geq 1}=0}=F(t).
   \]

 Let us consider some genus-zero correlation functions on the small phase space.
  The metric on the space of primary fields is defined by
 \[
\lan P\op_{\alpha}\op_{\beta}\ran=\eta_{\alpha\beta}.
 \]
When $\eta_{\alpha\beta}$ is independent of the couplings we call it the flat metric and
  the couplings $T^\alpha$ the flat coordinates. In fact, the three-point functions
   in the small phase space can be expressed as
\[
\lan \op_{\alpha}\op_{\beta}\op_{\gamma}\ran=\frac{\pa F}
 {\pa T^{\alpha}\pa T^{\beta}\pa T^{\gamma}} \equiv c_{\alpha\beta\gamma}
  \]
which provide the structure constants of the commutative and associative algebra
$\op_{\alpha}\op_{\beta}=c_{\alpha\beta}^\gamma \op_\gamma$.
 The associativity of $c_{\alpha \beta}^{\ga}$, i.e.,
$c_{\alpha \beta}^{\mu}c_{\mu \ga}^{\sigma}=c_{\alpha \ga}^{\mu}c_{\mu \beta}^{\sigma}$,
 will give the WDVV equations (\ref{WDVV}).

  Since the Lax hierarchy under consideration is a three-variable theory,
  thus only three primary fields
  $\{\op_1=\op_P\equiv P, \op_2=\op_Q\equiv Q, \op_3=\op_R\equiv R \}$
    are involved in the TFT formulation and we shall identify
 \[
    v^1|_{T^{\alpha,n\geq 1}=0}=T^R,\qquad v^2|_{T^{\alpha,n\geq 1}=0}=T^P+\frac{1}{4}(T^R)^2,
    \qquad v^3|_{T^{\alpha,n\geq 1}=0}=T^Q
 \]
   on the small phase space and the Lax operator   in small phase space is written as
   \[
   L(z)=z^2+T^Rz+T^P+\frac{1}{4}(T^R)^2+T^Qz^{-1}
 \]
    which can be viewed as a rational superpotential in LG formulation of TFT
     \cite{V,DVV}.
     According to the LG theory, the primary fields corresponding to $T^\alpha$
      are defined by
  \[
 \op_P(z)=\frac{\pa L(z)}{\pa T^P}=1,\qquad
 \op_Q(z)=\frac{\pa L(z)}{\pa T^Q}=z^{-1},\qquad
 \op_R(z)=\frac{\pa L(z)}{\pa T^Q}=z+\frac{T^R}{2}
  \]
which can be used to compute the three-point correlation functions through the
residue formula \cite{V,DVV}:
\be
c_{\alpha\beta\gamma}=\mbox{Res}_{L'=0}\left[
 \frac{\op_\alpha(z)\op_\beta(z)\op_\gamma(z)}{\pa_zL(z)}\right].
 \label{3pt}
\ee
 It is easy to show that (\ref{3pt}) reproduces the previous $c_{\alpha\beta\gamma}$ and $F$ on
  the small phase space. In particular, the non-zero components of the flat metric on the
   space of primary fields are given by $
\eta_{PP}=\eta_{QR}=\eta_{RQ}=1/2,~ \eta^{PP}=\eta^{QR}=\eta^{RQ}=2$
as we obtained previously. Now we define the fundamental correlation functions
 $\langle {\mathcal O}_\alpha{\mathcal O}_\beta\rangle=\pa^2F/\pa T^\alpha\pa T^\beta$
 as follows:
 \bean
 \langle PP\rangle=\frac{T^P}{2},\quad  \langle PQ\rangle&=&\frac{T^R}{2},\quad
  \langle PR\rangle=\frac{T^Q}{2},\\
 \langle QQ\rangle=\log T^Q,\quad \langle QR\rangle&=&\frac{T^P}{2}+\frac{(T^R)^2}{8},\quad
 \langle RR\rangle=\frac{T^QT^R}{4},
 \eean
    Although these two-point correlation functions are defined
    on the small phase space, however, it has been shown \cite{DW} that they can be defined
     on the full phase space through the variables $v^1$, $v^2$ and $v^3$ in which the
      gravitational couplings $T^{\alpha,n}$ do not vanish.
    Hence one can write down these genus-zero two-point functions on the full phase
     space and obtain the following constitutive relations:
 \bea
 \langle PP\rangle&=&\frac{v^2}{2}-\frac{(v^1)^2}{8},\quad
  \langle PQ\rangle=\frac{v^1}{2},\quad  \langle PR\rangle=\frac{v^3}{2},\no\\
 \langle QQ\rangle&=&\log v^3,\quad \langle QR\rangle=\frac{v^2}{2},\quad
 \langle RR\rangle=\frac{v^1v^3}{4},
 \label{consti}
 \eea
which will be important to provide the connection between the generalized Benney hierarchy
 and its associated  TFT.

  Based on the constitutive relations (\ref{consti}),
  we can identify the gravitational flows for $v^\alpha$ in the full phase space
  as the integrable flows by taking into account the genus-zero topological recursion
   relation \cite{W2,DW}:
\be
\langle \sigma_n(\op_\alpha)AB\rangle=\sum_{\beta,\gamma=P,Q,R}n\langle \sigma_{n-1}(\op_\alpha)
 \op_\beta\rangle \eta^{\beta\gamma}\langle \op_\gamma AB\rangle, \qquad \alpha=P,Q,R.
 \label{recursion}
\ee
 For example, setting $n=1,~\op_\alpha=P$ and $A=P, B=Q$ then
 \bean
\frac{\pa v^1}{\pa T^{P,1}} &=&2\langle \sigma_1(P)PQ\rangle\\
 &=&4[\langle PP\rangle\langle PPQ\rangle+\langle PQ\rangle\langle RPQ\rangle+
 \langle PR\rangle\langle QPQ\rangle\\
 &=&4[ \langle PP\rangle\lan PQ\ran'+\lan PQ\ran\lan RQ\ran'+\lan PR\ran\lan QQ\ran']\\
 &=&\left[v^1v^2-\frac{1}{12}(v^1)^3+2v^3\right]_X
 \eean
where we denote $f'=\pa f/\pa T^P=\pa f/\pa X$ . Similarly, taking $\op_\alpha=R$ we have
\[
\frac{\pa v^1}{\pa T^{R,1}}=2\lan \sigma_1(R)PQ\ran=\left[v^1v^3+\frac{1}{2}(v^2)^2\right]_X.
\]
On the other hand, taking $n=2$ we get
 \bean
 \frac{\pa v^1}{\pa T^{P,2}} &=&2\lan \sigma_2(P)PQ\ran\\
 &=&\left[\frac{1}{80}(v^1)^5-\frac{1}{6}(v^1)^3v^2+(v^1)^2v^3+v^1(v^2)^2+4v^2v^3\right]_X,\\
 \frac{\pa v^1}{\pa T^{R,2}} &=&2\lan \sigma_2(R)PQ\ran \\
 &=&\left[\frac{1}{3}(v^2)^3+2v^1v^2v^3+(v^3)^2\right]_X.
  \eean
 etc. We present other flows in Appendix A.

 Next we turn to the $T^{Q,n}$-flows. Choosing $\op_\alpha=Q$ and using the topological
 recursion relation (\ref{recursion}), we obtain
 \bean
\frac{\pa v^1}{\pa T^{Q,1}} &=&\left[\frac{1}{2}(v^1)^2+2v^2\log v^3\right]_X,\no\\
\frac{\pa v^1}{\pa T^{Q,2}} &=& \left[-\frac{1}{12}(v^1)^4+v^2(v^1)^2+2(v^2)^2\log v^3+
4v^1v^3(\log v^3-1)\right]_X.
 \eean
Comparing with the Lax equations (\ref{laxeq}) and (\ref{logflow})
 it turns out that the above recursion relations
can be recasted into the Lax form in terms of coupling times $T^{\alpha,n}$ as
 \bea
  \frac{\pa L}{\pa T^{P,n}}&=&\frac{2^nn!}{(2n+1)!!}\{L^{n+1/2}_{\ge 1}, L\},\no\\
 \frac{\pa L}{\pa T^{Q,n}}&=&\{[L^n(\log L-c_n)]_{\ge 1}, L\}.\no\\
 \frac{\pa L}{\pa T^{R,n}}&=&\frac{1}{2n+2}\{L^{n+1}_{\ge 1}, L\}
 \label{pqrflow}
 \eea
 with the identification
\be
T_{2n-1}=\frac{2^{n-1}(n-1)!}{(2n-1)!!}T^{P,n-1},\quad T_{2n}= \frac{1}{2n}T^{R,n-1},
\quad \bar{T}_n=\frac{1}{2}T^{Q,n}.
\label{identify}
\ee
 On the other hand, the associated commuting Hamiltonian flows with respect
  to the bi-Hamiltonian structures are thus given by
 \bea
\frac{\pa {\mathbf{v}}}{\pa T^{P,n}}&=&\{H_{P,n+1}, {\mathbf{v}}\}_1=
\frac{2n}{2n+1}\{H_{P,n},{\mathbf{v}}\}_2,
 \qquad H_{P,n}=\frac{2^n(n-1)!}{(2n+1)!!}\int \res L^{n+1/2}\no\\
 \frac{\pa {\mathbf{v}}}{\pa T^{R,n}}&=&\{H_{R,n+1}, {\mathbf{v}}\}_1=
\frac{n}{n+1}\{H_{R,n},{\mathbf{v}}\}_2,
 \qquad H_{R,n}=\frac{1}{2n(n+1)}\int \res L^{n+1}\no\\
 \frac{\pa {\mathbf{v}}}{\pa T^{Q,n}}&=&\{H_{Q,n+1},
{\mathbf{v}}\}_1=\{H_{Q,n}, {\mathbf{v}}\}_2,\qquad H_{Q,n}=
 \frac{3}{2n}\int \res [L^n(\log L-c_n)]
 \label{hflow}
 \eea
 where $n=0,1,2,\cdots$.

Furthermore, using (\ref{flat}), (\ref{consti}) and  (\ref{pqrflow}) the  flows for $t^\alpha$
are
 \bea
\frac{\pa t^1}{\pa T^{\alpha,n}}&=& 2\lan\sigma_n(\op_\alpha)PP\ran=(R^{(1)}_{\alpha,n})',\no\\
 \frac{\pa t^2}{\pa T^{\alpha,n}}&=& 2\lan\sigma_n(\op_\alpha)PR\ran=(R^{(2)}_{\alpha,n})',\no\\
 \frac{\pa t^3}{\pa T^{\alpha,n}}&=& 2\lan\sigma_n(\op_\alpha)PQ\ran=(R^{(3)}_{\alpha,n})'
 \label{tflow}
 \eea
where $R^{(\beta)}_{\alpha,n}$ are the analogues of the Gelfand-Dickey potentials \cite{Di} of the
 KP hierarchy, given by
 \bea
  R^{(1)}_{P,n}&=&2\lan\sigma_n(P)P\ran=\frac{2^{n+1}n!}{(2n+1)!!}\res(L^{n+1/2}),\no\\
  R^{(1)}_{Q,n}&=&2\lan\sigma_n(Q)P\ran=2\res[L^n(\log L-c_n)],\no\\
  R^{(1)}_{R,n}&=&2\lan\sigma_n(R)P\ran=\frac{1}{n+1}\res(L^{n+1}),\no\\
  R^{(2)}_{P,n}&=&2\lan\sigma_n(P)R\ran=\frac{2^nn!}{(2n+1)!!}
  [2(L^{n+1/2})_{-2}+v^1\res(L^{n+1/2})],\no\\
  R^{(2)}_{Q,n}&=&2\lan\sigma_n(Q)R\ran=[2(L^n(\log L-c_n))_{-2}+v^1\res(L^n(\log L-c_n))],\no\\
  R^{(2)}_{R,n}&=&2\lan\sigma_n(R)R\ran=\frac{1}{2n+2}[2(L^{n+1})_{-2}+v_1\res(L^{n+1})],\no\\
   R^{(3)}_{P,n}&=&2\lan\sigma_n(P)Q\ran=\frac{2^{n+1}n!}{(2n+1)!!}(L^{n+1/2})_0,\no\\
  R^{(3)}_{Q,n}&=&2\lan\sigma_n(Q)Q\ran=2[L^n(\log L-c_n)]_0,\no\\
  R^{(3)}_{R,n}&=&2\lan\sigma_n(R)Q\ran=\frac{1}{n+1}(L^{n+1})_0.
 \label{gdpot}
 \eea
 In particular, the two-point correlators involving the first and the second
  descendants are also presented in  Appendix A.

 Finally, we would like to remark that the first three equations of (\ref{gdpot}) can
  be integrated to yield
 \bean
 \lan\sigma_n(P)\ran&=&\frac{2^{n+1}n!}{(2n+3)!!}\res L^{n+3/2},\\
 \lan\sigma_n(Q)\ran&=&\frac{1}{n+1}\res[L^{n+1}(\log L-c_{n+1})],\\
\lan\sigma_n(R)\ran&=&\frac{1}{2(n+1)(n+2)}\res L^{n+2}
 \eean
which represent the one-point functions of gravitational descendants
 at genus zero in the LG description.

\section{Twistor construction for the string equation}

After coupling to gravity, the dynamical variables $v^\alpha$ (or flat coordinates $t^\alpha$)
become functions on the full phase space that is parametrized by $T^{\alpha, n}$. Then
$v^\alpha$ (or $t^\alpha$) satisfy not only the Lax flows but also the string equation
which characterizes the corresponding additional symmetries or $W$-constraint associated with the
generalized Benney hierarchy.  Since the Lax formulation in sec. 3 for the generalized
Benney hierarchy is  similar to that of the dToda type hierarchy  \cite{TT,TT1}.
 This motivates us to   reproduce the generalized Benney equations by imposing constraints on
the Lax operators  and the associated Orlov operators of the dmKP hierarchy \cite{CT1}
 through the twistor data.  We shall follow closely that of \cite{Kh} to show that the constraints
imposing on the twistor data  implies the string equation of the generalized Benney
hierarchy. Let us consider two Lax operators $ \mu$ and $\tilde{ \mu}$ with
the following
 Laurent expansions
 ($T^1=X$):
  \be
   \mu = p+\sum_{n=0}^{\infty} v^{n}( T,  \tilde{ T}) p^{-n},\qquad
    {\tilde{ \mu}}^{-1} = \tilde {v}_{0}(T, \tilde {T})
p^{-1}+\sum_{n=0}^{\infty} \tilde {v}_{n+1}(T, \tilde {T})p^n
 \label{pole}
 \ee
  which satisfy the commuting Lax flows
   \bea
    \frac{\pa \mu}{\pa T_n} &=& \{B_n, \mu \}, \qquad  \frac{\pa \mu}{\pa
\tilde {T}_n}
     = \{\tilde {B}_n, \mu \}, \no \\
      \frac{\pa \tilde {\mu}}{\pa T_n} &=& \{B_n, \tilde {\mu} \}, \qquad
       \frac{\pa \tilde {\mu}}{\pa \tilde {T}_n} = \{\tilde {B}_n, \tilde {\mu} \},
 \label{bar1}
 \eea
  where
\[
B_n\equiv (\mu^n)_{\geq 1}, \qquad \tilde {B}_n\equiv ({\tilde {\mu}}^{-n})_{\leq 0},
\quad (n=1,2,3 \cdots)
 \]
  Next, we consider the Orlov operators corresponding to dmKP
 \be
  M = \sum_{n=1}^{\infty}nT_n
\mu^{n-1}+\sum_{n=1}^{\infty}w_n(T,\tilde {T})\mu^{-n},\qquad
 \tilde {M} =- \sum_{n=1}^{\infty} n \tilde{T}_n {\tilde
{\mu}}^{-n-1}+X+\sum_{n=1}^{\infty}
 \tilde {w}_n (T,\tilde {T}){\tilde {\mu}}^{n}
   \ee
  with constraint
    \bea
\{\mu, M\}=1, \qquad \{\tilde {\mu}, \tilde{ M} \}=1.
  \label{unit}
 \eea
  In fact, the coefficient functions $w_n$ and $\tilde {w}_n$ in the
Orlov operators are defined by the above canonical relations and the following flow
equations
 \bea
  \frac{\pa M}{\pa T_n} &=& \{B_n, M \},
\qquad  \frac{\pa M}{\pa \tilde{T}_n} = \{\tilde{B}_n, M \},\no \\
 \frac{\pa \tilde{M}}{\pa T_n} &=& \{B_n, \tilde{M} \}, \qquad \frac{\pa
\tilde {M}}
 {\pa \tilde{T}_n}
 = \{\tilde{B}_n, \tilde{M} \}.
  \label{bar2}
   \eea
   Inspired by the twistor construction (or Riemann-Hilbert problem) for
the solution structure of the dToda hierarchy \cite{TT,TT1}, we now give the
twistor construction for the generalized Benney hierarchy.\\
{\sc Theorem 1} \cite{CT3}.  Let  $f(p,X),g(p,X), \tilde {f}(p,X), \tilde
{g}(p,X)$ be functions satisfying
 \bea
  \{f(p,X), g(p,X)\}=1, \qquad \{\tilde {f}(p,X), \tilde {g}(p,X)\}=1.
 \label{pois}
  \eea
   Then the functional equations
    \bea
     f(\mu, M)=\tilde {f}(\tilde {\mu}, \tilde {M}),
    \qquad g(\mu, M)=\tilde {g} (\tilde {\mu}, \tilde {M})
     \label{fun}
\eea
 provide a solution of (\ref{bar1}) and (\ref{bar2}).
  We call the pairs $(f,g)$ and $(\tilde {f}, \tilde {g})$ the twistor
data of the solution.

To apply the above Theorem to the present case, we have
 to impose  the following constraint
\begin{equation}
L=\mu^2={\tilde{\mu}}^{-1}, \label{tw1}
\end{equation}
that is, $f(p,X)=p^2$ and $\tilde {f}(p,X)=p^{-1}$.
  As a result, the time variables $\tilde {T}_n$
  can be eliminated via the following identification:
\[
 \tilde {T}_n=-T_{2n}.
 \]
From (\ref{pois}), the twistor data $g(p,X)$ and $\tilde {g}(p,X)$ can be
 assumed to be in the following form
\be
g(p,X)= \frac{X}{2p}-\sum_{n=1}^{\infty} nT_{2n} p^{2n-2},\qquad
 \tilde {g}(p,X) =-Xp^2,
  \label{gfunc}
\ee
where the second part of $g(p,X)$ is responsible for the string equations
(see below).
 By the Theorem 1 and equation (\ref{gfunc}), we get the following constraint
  for the Orlov operators:
\begin{equation}
   \frac{M\mu^{-1}}{2} -\sum_{n=1}^{\infty} n T_{2n} \mu^{2n-2}=-{\tilde
{\mu}}^2 \tilde {M}.
 \label{tw2}
 \end{equation}
  It's the above constraint that leads to the string equations.
So far, the twistor construction only involves the Lax flows (\ref{laxeq}).
 To obtain the string equations associated with the underlying TFT described
 in sec. 3, we have to properly extend the standard Orlov operator
 to include the additional hierarchy equations (\ref{logflow}). Namely,
   it's necessary to introduce the flows generated by the logarithmic
operator
\[
\bar {B_n}=[L^n(\log L-c_n)]_{\geq 1}.
 \]
   Let $ \bar {T_n} $ be the time variables of additional flows generated
by $\bar {B_n}$
 and then the Orlov operator $M$ should be deformed  by these new flows to $M'$
so that
 \begin{equation}
  \frac{\pa M'}{\pa \bar {T_n}}=2 \{\bar {B_n} ,M' \}.
   \label{tw3}
   \end{equation}
   To construct the modified Orlov operator $M'$, it is convenient to
using the dressing method \cite{TT1}. Let us first express the original Lax operator
$\mu$ and its conjugate Orlov operator $M$ in dressing form
\begin{equation}
 \mu = e^{\mbox{ad}\Th}(p),\qquad
M=e^{\mbox{ad}\Th}\left(\sum_{n=1}^{\infty} nT_n p^{n-1}\right)
    \label{dress1}
\end{equation}
    where
    \[
 e^{\mbox{ad}f}(g)=g+\{f,g\}+\frac{1}{2!} \{f, \{f,g\} \}+\cdots.
    \]
     One can understand that this is the canonical transformation
generated by $\Th(T,\bar{T},p)$
     (for the $\bar{T}$-dependence, see below) and its flow equations
(Sato equations)
      can be written as \cite{TT1}
\begin{equation}
   \nabla_{T_n,\Th}\Th=B_n-e^{\mbox{ad}\Th}(p^n)
    \label{sato}
\end{equation}
    where
    \[
    \nabla_{T_n,\Th}f \equiv
\sum_{k=0}^{\infty}\frac{1}{(k+1)!}(\mbox{ad}\Th)^k
    \left(\frac{\pa f}{\pa T_n}\right).
    \]
    It is easy to show that (\ref{sato}) together with (\ref{dress1})
     implies the flow equations (\ref{bar2}).
In contrast with equation (\ref{sato}), the $\bar {T_n}$ flows for $\Th$
are given by
\[
 \nabla_{\bar{T}_n,\Th}\Th=2\bar{B}_n-2e^{\mbox{ad}\Th}[p^{2n}(\log
p^2-c_n)]
 \]
  and a similar argument reaches the modified Orlov operator
 \begin{eqnarray}
  M'&=& e^{\mbox{ad}\Th} \left(\sum_{n=1}^{\infty}n T_{n} p^{n-1}+4
\sum_{n=1}^{\infty} n \bar {T_n}p^{2n-1}(\log p^2-c_{n-1}) \right )\no\\
 &=& \sum_{n=1}^{\infty}2n T_{2n}
{\mu}^{2n-1}+\sum_{n=0}^{\infty}(2n+1) T_{2n+1}
{\mu}^{2n}+\sum_{n=1}^{\infty} w_n {\mu}^{-n}\no \\
  &&+  4 \sum_{n=1}^{\infty} n \bar {T}_n{\mu}^{2n-1}(\log \mu^2 -c_{n-1}),
   \label{modm}
 \end{eqnarray}
 which  satisfies the additional flow equations (\ref{tw3}).

 Now we are in the position to derive the string equation.
Let us decompose $(M'\mu^{-1}/2-\sum_{n=1}^{\infty} n T_{2n}{\mu}^{2n-2})$ into
the positive power part
 $(M'\mu^{-1}/2 -\sum_{n=1}^{\infty} n T_{2n}{\mu}^{2n-2})_{\geq
1}$ and the non-positive power part   $(M'\mu^{-1}/2-\sum_{n=1}^{\infty} n
T_{2n}{\mu}^{2n-2})_{\leq 0}$ by using equations (\ref{tw2}) and
(\ref{modm}).
  It turns out  that
 \bean
 \left(\frac{M'\mu^{-1}}{2} -\sum_{n=1}^{\infty} n T_{2n}
{\mu}^{2n-2} \right)_{\geq 1}
    &=&  \sum_{n=1}^{\infty} \frac{(2n+1)}{2}
T_{2n+1} B_{2n-1} + 2\sum_{n=1}^{\infty}n \bar{T_n} \bar{B}_{n-1}, \\
 \left(\frac{M'\mu^{-1}}{2} -\sum_{n=1}^{\infty}n
T_{2n} {\mu}^{2n-2} \right)_{\leq 0}&=& -\left( {\tilde {\mu}}^2 \tilde
{M} \right)_{\leq 0}=-\left( \sum_{n=1}^{\infty} n T_{2n} {\mu}^{2(n-1)} \right)_{\leq 0}
 \eean
 where, by the definition (\ref{pole}), we have used the fact that
  $(\mu^{-2n})_{\geq 1}=(\tilde{\mu}^{n})_{\leq 0}=0$ for $n\geq 1$.
   Hence
\[
     g(\mu,M') = \sum_{n=1}^{\infty} \frac{(2n+1)}{2} T_{2n+1}
B_{2n-1}+ 2 \sum_{n=1}^{\infty}n \bar{T_n} \bar B_{n-1}-\left(
\sum_{n=1}^{\infty}
     n T_{2n} {\mu}^{2(n-1)} \right)_{\leq 0}
\]
 which together with the canonical commutation relation $\{g(\mu, M'), L
\}=-1$ implies
   \[ \sum_{n=2}^{\infty} n T_{2n} \frac{\pa L}{\pa
T_{2n-2}}+\sum_{n=1}^{\infty} \frac{(2n+1)}{2} T_{2n+1} \frac{\pa L}{\pa
T_{2n-1}}+\sum_{n=1}^{\infty} n \bar{T_n} \frac{\pa
L}{\pa {\bar T}_{n-1}}=-1
 \]
 where the Lax flows (\ref{laxeq}) and (\ref{logflow}) have been used.
  Therefore by (\ref{identify}) we obtain that
\[ \sum_{n=1}^{\infty} n T_{R,n} \frac{\pa L}{\pa
T^{R,n-1}}+\sum_{n=1}^{\infty}n T^{P,n} \frac{\pa L}{\pa
T^{P,n-1}}+\sum_{n=1}^{\infty} n T^{Q,n}
 \frac{\pa  L}{\pa T^{Q,n-1}}=-1.
 \]
  After shifting $T^{P,1} \to T^{P,1}-1$, we get
\[ \sum_{n=1}^{\infty} n T^{R,n} \frac{\pa L}{\pa
T^{R,n-1}}+\sum_{n=1}^{\infty}n T^{P,n} \frac{\pa L}{\pa T^{P,n-1}}+
 \sum_{n=1}^{\infty} n T^{Q,n} \frac{\pa  L}{\pa T^{Q,n-1}}
+1=  \frac{\pa L}{\pa T^{P,0}}.
 \]
Comparing the both-hand sides of the above equation in $p$ and taking
into account the constitutive relations (\ref{consti}) we obtain the following genus-zero
string equation
 \bean
 t^1(T)&=&T^P+2\sum_{n=1}^{\infty}\sum_{\alpha=P,Q,R}
 nT^{\alpha,n}\lan \sigma_{n-1}({\mathcal O}_\alpha)P\ran,\\
 t^2(T)&=&T^Q+2\sum_{n=1}^{\infty}\sum_{\alpha=P,Q,R}
 nT^{\alpha,n}\lan \sigma_{n-1}({\mathcal O}_\alpha)R\ran,\\
 t^3(T)&=&T^R+2\sum_{n=1}^{\infty}\sum_{\alpha=P,Q,R}
 nT^{\alpha,n}\lan \sigma_{n-1}({\mathcal O}_\alpha)Q\ran
 \eean
 which gives a special solution of the generalized Benney hierarchy and is used
 to characterize the free energy in the full phase space. As a consistent check,
differentiating $t^{\alpha}$ to $T^{\beta,n}$ and restricting to the small phase space
we recover the flows (\ref{tflow}) as expected.

\section{quantization of the Poisson brackets}

So far, the TFT approach to the generalized Benney hierarchy is restricted to the
genus-zero sector. In general it is not easy to promote the Benney system
to higher genera through the TFT formulation due to the complexity of topological
 recursion relations. Nevertheless there are at least two approaches enable us to
  get consistent genus-one extensions for some integrable hierarchies.
  One of them is based on the  large-$N$ matrix expansion of integrable
  hierarchies developed by Eguchi and Yang \cite{EY} and the other,
  the theory of Frobenius manifolds by DZ \cite{DZ}.
In this section, we shall compute the genus-one correction to the
 bi-Hamiltonian structure  of the generalized Benney hierarchy by using
the latter approach. Roughly speaking, the  DZ approach consists of the following two main steps:
\begin{itemize}
\item introducing slow spatial and time variables scaling
\[
 T^{\alpha, n} \to \ep T^{\alpha,n},
 \label{scaling}
 \]
\item changing the full free energy as
\[ {\mathcal F\/} \to \sum_{g=0}^{\infty} \ep^{2g-2} {\mathcal F\/}_g,
 \]
\end{itemize}
where $\ep$ is the parameter of genus expansion.
 As a result, all of the corrections  become series in $\ep$.

  To get a unambiguous genus-one correction of the Hamiltonian flows
(\ref{hflow}) it is convenient to expand the flat coordinates up to the $\ep^2$ order as
  \[
 t_\alpha=t_\alpha^{(0)}+\ep^2t_\alpha^{(1)}+O(\ep^4),\qquad
t_\alpha=\eta_{\alpha\beta}t^\beta
   \]
 where $t_\alpha^{(0)}$ is the ordinary flat coordinates $t_\alpha$ satisfying
  (\ref{tflow}) and $t_\alpha^{(1)}$ is the genus-one correction defined by
 $t_\alpha^{(1)}=\pa^2{\mathcal F\/}_1(T)/\pa T^\alpha\pa X$.
Then there exists a unique  hierarchy flows of the form \cite{DZ}
 \be
  \frac{\pa t^\alpha}{\pa T^{\beta,n}}=\{t^\alpha(X),
H_{\beta,n+1}\}_1=\{t^\alpha(X), H_{\beta,n}\}_2
  \label{1hflow}
 \ee
with
 \bean
 \{t^\alpha (X), t^\beta (Y)\}_i&=&\{t^\alpha (X), t^\beta (Y)\}^{(0)}_i+
 \ep^2\{t^\alpha (X), t^\beta (Y)\}^{(1)}_i+O(\ep^4),\qquad i=1,2\no\\
 H_{\alpha,n}&=&H_{\alpha,n}^{(0)}+\ep^2H_{\alpha,n}^{(1)}+O(\ep^4)
\eean
   That means under such correction the Poisson brackets $J_1$ and $J_2$
and the Hamiltonians
  will receive corrections up to $\ep^2$ such that the Hamiltonian flows
   (\ref{1hflow}) still commute with each other.
In \cite{DZ}, the genus-one part of the free energy can be expressed as
 \bea
  {\mathcal F\/}_1(T)=\left[\frac{1}{24}
\log \det M^{\alpha}_\beta(t, \pa_X t)+ G(t)\right]_{t=t(T)},
 \label{genus}
 \eea
  where the matrix $M^{\alpha}_\beta$ is given by
   \[
    M^{\alpha}_\beta(t, \pa_X t)= c^{\alpha}_{\beta\ga}(t) \pa_X t^{\ga}=
     \left(
 \begin{array}{ccc} t_X^1 & t_X^3& t_X^2 \\
   t_X^2 &t_X^1+\frac{t^3 t^3_X}{2} & \frac{t^3 t^2_X+ t^3_X t^2}{2} \\
   t_X^3 & \frac{2t_X^2}{t^2}& t_X^1+\frac{t^3 t^3_X}{2}
  \end{array}
    \right)
 \]
  and $G(t^2,t^3)$ is the associated $G$-function satisfying the Getzler's equation
(\ref{geq}) which are over-determined and some of them
are redundant. In general, they can be solved explicitly using canonical
coordinates for an arbitrary semi-simple Frobenius manifold \cite{Du2}.
Here we notice that $c_{22}^3=2/t^2, $ and $c_{22}^2=c_{22}^1=0$ and
  substitute them into  the coefficient of $(z_2)^4$ of the Getzler
equations (\ref{geq}) to get the following ordinary differential equation
\[
 \frac{\pa^2 G}{\pa (t^2)^2}+ \frac{1}{2t^2}\frac{\pa G}{\pa t^2}-
\frac{1}{24(t^2)^2} =0
\]
which together with the quasi-homogeneous property ${\mathcal L\/}_EG=-1/8$
(c.f. Theorem 3 in \cite{DZ}) implies (up to a constant)
\begin{equation}
G(t_2,t_3)= -\frac{\log t^2}{12} .
 \label{gfn2}
\end{equation}
We remark that the above $G$-function is the same as the one
 of the Benney hierarchy with two primary fields \cite{CT3}.

Thus a simple computation yields
 \begin{eqnarray*}
   {\mathcal F\/}_1 &=&\frac{1}{24} \log
\left[t^2(t^1_X)^3+2(t^2_X)^3+\frac{(t^2)^2(t^3_X)^3}{2}
+t^2t^3(t_X^1)^2 t^3_X + \frac{t^2 (t^3_X)^2(t^3)^2 t^2t^1_X}{4}\right.  \\
 &&\left.-t^3 t_X^1 (t^2_X)^2 - 3t^2 t^1_X t^2_X t^3_X -\frac{t^2t^3t^2_X
(t^3_X)^2}{2}\right]-\frac{\log t^2}{8}.
\end{eqnarray*}
Using $c^\alpha_{\beta\gamma}$ and ${\mathcal F\/}_1$ and consulting the procedure
 developed by DZ (c.f Theorems 1 and 2 in \cite{DZ}) it turns out that
 the genus-one corrections of the Poisson brackets, in terms of flat coordinates $t^\alpha$,
  are given by
    \bean
  \{t^1(X),t^2(Y)\}_1&=&\{t^2(X),t^2(Y)\}_1=\{t^3(X),t^3(Y)\}_1=O(\ep^4),\no\\
  \{t^1(X),t^1(Y)\}_1&=&2\de'(X-Y)+O(\ep^4),\no\\
\{t^1(X),t^3(Y)\}_1&=&-\frac{2\ep^2}{3}\left[\frac{\de'''(X-Y)}{t^2}-
\frac{2t^2_X\de''(X-Y)}{(t^2)^2}+\frac{2(t^2_X)^2\de'(X-Y)}{(t^2)^3}\right.\no\\
&&\left.-\frac{t^2_{XX}\de'(X-Y)}{(t^2)^2}\right]+O(\ep^4),\no\\
    \{t^2(X),t^3(Y)\}_1&=&2\de'(X-Y)-\frac{\ep^2}{3}\left[\frac{t^3\de'''(X-Y)}{t^2}
    -\frac{2t^3t^2_X\de''(X-Y)}{(t^2)^2}+\frac{t^3_X\de''(X-Y)}{t^2}\right.\no\\
    &&\left.-\frac{t^3_Xt^2_X\de'(X-Y)}{(t^2)^2}-\frac{t^3t^2_{XX}\de'(X-Y)}{(t^2)^2}
    +\frac{2t^3(t^2_X)^2\de'(X-Y)}{(t^2)^3}\right]+O(\ep^4),
   \label{pb1}
 \eean
 and
 \bean
\{t^1(X),t^1(Y)\}_2&=&t^1_X\de(X-Y)+2t^1\de'(X-Y)+\frac{\ep^2\de'''(X-Y)}{2}
+O(\ep^4),\no\\
  \{t^1(X),t^2(Y)\}_2&=&2t^2_X\de(X-Y)+3t^2\de'(X-Y)+O(\ep^4),\no\\
  \{t^1(X),t^3(Y)\}_2&=&t^3\de'(X-Y)+\ep^2\left[-\frac{4t^1+(t^3)^2}{6t^2}\de'''(X-Y)\right.\no\\
  &&\left.+\frac{4t^1t^2_X+(t^3)^2t^2_X-4t^2t^1_X-2t^2t^3t^3_X}{3(t^2)^2}\de''(X-Y)\right.\no\\
&&\left.+\left(\frac{4t^1t^2_X+(t^3)^2t^2_X-4t^2t^1_X-2t^2t^3t^3_X}{6(t^2)^2}\right)_X\de'(X-Y)\right]
  +O(\ep^4),\no\\
  \{t^2(X),t^2(Y)\}_2&=&2t^2t^3\de'(X-Y)+(t^2t^3)_X\de(X-Y)\no\\
  &&+\ep^2\left[\frac{1}{3}\left(2t^1_{XX}+(t^3_X)^2+t^3t^3_{XX}
  -\frac{2t^2_Xt^1_X}{t^2}-\frac{t^2_Xt^3t^3_X}{t^2}\right)\de'(X-Y)\right.\no\\
  &&\left.+\frac{1}{6}\left(2t^1_{XXX}+3t^3_Xt^3_{XX}+t^3t^3_{XXX}-
  \left(\frac{t^3t^3_Xt^2_X+2t^1_Xt^2_X}{t^2}\right)_X\right)\de(X-Y)\right]+O(\ep^4),\no\\
\{t^2(X),t^3(Y)\}_2&=&\left(2t^1+\frac{(t^3)^2}{2}\right)\de'(X-Y)+
\ep^2\left[\left(1-\frac{t^1t^3}{3t^2}-\frac{(t^3)^3}{12t^2}\right)\de'''(X-Y)\right.\no\\
&&+\left(\frac{8t^3t^1t^2_X+2(t^3)^3t^2_X-8t^2t^3t^1_X-5t^2(t^3)^2t^3_X-4t^2t^1t^3_X-4t^2t^2_X}
{12(t^2)^2}\right)\de''(X-Y)\no\\
  &&+\left(\left(\frac{t^3t^2_X}{3(t^2)^2}\right)_Xt^1-\frac{(t^3)^3(t^2_X)^2}{6(t^2)^3}+
\frac{5(t^3)^2t^2_Xt^3_X}{12(t^2)^2}-\frac{t^3(t^3_X)^2}{3t^2}-\frac{(t^3)^2t^3_{XX}}{6t^2}\right.\no\\
&&\left.\left.-\frac{t^3_Xt^1_X}{3t^2}+\frac{(t^3)^3t^2_{XX}}{12(t^2)^2}+\frac{2t^3t^2_Xt^1_X}{3(t^2)^2}
-\frac{(t^2_X)^2}{6(t^2)^2}-\frac{t^3t^1_{XX}}{3t^2}\right)\de'(X-Y)\right]+O(\ep^4),\no\\
\{t^3(X),t^3(Y)\}_2&=&6\de'(X-Y)-\ep^2\left[\frac{2t^3}{3t^2}\de'''(X-Y)+
\left(\frac{t^3}{t^2}\right)_X\de''(X-Y)\right.\no\\
&&\left.+\left(\frac{t^3}{3t^2}\right)_{XX}\de'(X-Y)\right]+O(\ep^4)
  \label{pb2}
 \eean
which still satisfy the relation $J_1^{ij}=\pa J_2^{ij}/\pa t^1$.
For the genus-one corrections to the Hamiltonians $H_{\alpha,n}$ we collect
some of them in Appendex B.


\section{Concluding Remarks}
We have studied the generalized Benney hierarchy from the topological field
theory . First of all, we can construct the free energy from the
bi-Hamiltonian structure of the generalized Benney  hierarchy using the
flat pencil. Secondly, after introducing the flat coordinates, the
primary fields in the LG formulation of TFT can be defined.
Then the topological recursion relations between gravitational
descendants at genus-zero turn out to be the associated commuting Hamiltonian flows in
the generalized Benney hierarchy. Furthermore, we use the twistor
construction to obtain the string equation associated with the TFT, which
is a Galiean-type symmetry of the generalized Benney hierarchy and
characterizes the free energy in the full phase space. Finally, based on
the approach of DZ, we obtain the genus-one corrections of the
Poisson brackets. \\
\indent In spite of the results obtained, there are some interesting
issues deserving more investigations:
\begin{itemize}
\item In \cite{CT3}, it can be shown that the string coupling constant
$\ep$ of the Benney hierarchy will correspond to the effect of dispersion
of Kaup-Broer hierarchy \cite{OS,KO}. However, for the case of the generalized Benney
hierarchy, the computation is more involved and needs further
investigations.
\item The relations of free energies between dKP and dmKP are interesting.
 In \cite{AK}, the free energy of dKP is
constructed using generalized Gelfand-Dickey potential without referring
to  the bi-Hamiltonian structure of dKP. A similar approach for dmKP
should be developed and see what role is played by the Miura map between dKP and dmKP \cite{CT1}.
\item Recently, the associativity equations of infinite many primary fields
in dKP are obtained \cite{BMRWZ} by the dispersionless Hirota equations of
dKP\cite{CK}. Thus it is quite natural to ask whether the dmKP hierarchy and its reductions
 have the similar structures of associativity equations and what kind of infinite dimensional
  Frobenius manifold \cite{GT} they provide.
  \end{itemize}

{\bf Acknowledgements\/}\\
 JHC thanks Y. Kodama for helpful discussions.
 MHT thanks the National Science Council of Taiwan
 (Grant No. NSC 89-2112-M194-020) for support.
\newpage

\appendix
\section{Primary flows and two-point functions}
\subsection{Primary $T^{P,n}$-flows}
 \bean
 \frac{1}{2}\frac{\pa v^1}{\pa T^{P,0}}&=&\lan PQ\ran_X,\\
 \frac{1}{2}\frac{\pa v^2}{\pa T^{P,0}}&=&\lan QR\ran_X,\\
 \frac{1}{2}\frac{\pa v^3}{\pa T^{P,0}}&=&\lan PR\ran_X,\\
 \frac{1}{2}\frac{\pa v^1}{\pa T^{P,1}}&=&\lan \sigma_1(P)PQ\ran=
 \left[\frac{1}{2}v^1v^2-\frac{1}{24}(v^1)^3+v^3\right]_X,\\
 \frac{1}{2}\frac{\pa v^2}{\pa T^{P,1}}&=&\lan \sigma_1(P)QR\ran=
 \frac{1}{2}v^2v^2_X+\frac{1}{8}(v^1)^2v^2_X+v^1v^3_X+\frac{1}{2}v^1_Xv^3,\\
 \frac{1}{2}\frac{\pa v^3}{\pa T^{P,1}}&=&\lan \sigma_1(P)PR\ran=
 \left[\frac{1}{2}v^2v^3+\frac{1}{8}(v^1)^2v^3\right]_X,\\
  \frac{1}{2}\frac{\pa v^1}{\pa T^{P,2}}&=&\lan \sigma_2(P)PQ\ran=
 \left[\frac{1}{160}(v^1)^5-\frac{1}{12}(v^1)^3v^2+\frac{1}{2}(v^1)^2v^3+
 \frac{1}{2}v^1(v^2)^2+2v^2v^3\right]_X,\\
   \frac{1}{2}\frac{\pa v^2}{\pa T^{P,2}}&=&\lan \sigma_2(P)QR\ran=
   -\frac{1}{96}(v^1)^4v^2_X+\frac{1}{6}(v^1)^3v^3_X+\frac{1}{4}(v^1)^2v^2v^2_X+
   \frac{1}{4}(v^1)^2v^1_Xv^3+2v^1(v^2v^3)_X\\
   &&+\frac{1}{6}(v^2)^3_X+(v^3)^2_X+v^1_Xv^2v^3,\\
 \frac{1}{2}\frac{\pa v^3}{\pa T^{P,2}}&=&\lan
 \sigma_2(P)PR\ran=\left[-\frac{1}{96}(v^1)^4v^3+\frac{1}{4}(v^1)^2v^2v^3+
 v^1(v^3)^2+\frac{1}{2}(v^2)^2v^3\right]_X.
 \eean
From above, we can extract the following two-point functions:
 \bean
 \lan \sigma_1(P)P\ran&=&\frac{1}{64}(v^1)^4-\frac{1}{8}(v^1)^2v^2+\frac{1}{2}v^1v^3+
 \frac{1}{4}(v^2)^2,\\
 \lan \sigma_1(P)Q\ran&=&\frac{1}{2}v^1v^2-\frac{1}{24}(v^1)^3+v^3,\\
 \lan \sigma_1(P)R\ran&=&\frac{1}{2}v^2v^3+\frac{1}{8}(v^1)^2v^3,\\
 \lan \sigma_2(P)P\ran&=&-\frac{1}{384}(v^1)^6+\frac{1}{32}(v^1)^4v^2-\frac{1}{12}(v^1)^3v^3
 -\frac{1}{8}(v^1)^2(v^2)^2+v^1v^2v^3+\frac{1}{6}(v^2)^3+(v^3)^2,\\
 \lan \sigma_2(P)Q\ran&=&\frac{1}{160}(v^1)^5-\frac{1}{12}(v^1)^3v^2+\frac{1}{2}(v^1)^2v^3+
 \frac{1}{2}v^1(v^2)^2+2v^2v^3,\\
 \lan \sigma_2(P)R\ran&=&-\frac{1}{96}(v^1)^4v^3+\frac{1}{4}(v^1)^2v^2v^3+
 v^1(v^3)^2+\frac{1}{2}(v^2)^2v^3.
  \eean

\subsection{Primary $T^{Q,n}$-flows}
 \bean
 \frac{1}{2}\frac{\pa v^1}{\pa T^{Q,0}}&=&(\log(v^3))_X,\\
 \frac{1}{2}\frac{\pa v^2}{\pa T^{Q,0}}&=&\frac{1}{2}v^1_X+\frac{1}{2}v^1(\log (v^3))_X,\\
 \frac{1}{2}\frac{\pa v^3}{\pa T^{Q,0}}&=&\frac{1}{2}v^2_X,\\
 \frac{1}{2}\frac{\pa v^1}{\pa T^{Q,1}}&=&\lan \sigma_1(Q)PQ\ran=
 \left[\frac{1}{4}(v^1)^2+v^2\log v^3\right]_X,\\
 \frac{1}{2}\frac{\pa v^2}{\pa T^{Q,1}}&=&\lan \sigma_1(Q)QR\ran=
 \frac{1}{2}(v^1v^2)_X+v^3_X\log v^3+\frac{1}{2}v^1v^2(\log v^3)_X+
 \frac{1}{2}v^1v^2_X\log v^3,\\
 \frac{1}{2}\frac{\pa v^3}{\pa T^{Q,1}}&=&\lan \sigma_1(Q)PR\ran=
 \left[\frac{1}{4}(v^2)^2+\frac{1}{2}v^1v^3\log v^3\right]_X,\\
  \frac{1}{2}\frac{\pa v^1}{\pa T^{Q,2}}&=&\lan \sigma_2(Q)PQ\ran=
 \left[-\frac{1}{24}(v^1)^4+\frac{1}{2}v^2(v^1)^2+(v^2)^2\log v^3+2v^1v^3(\log v^3-1)\right]_X,\\
   \frac{1}{2}\frac{\pa v^2}{\pa T^{Q,2}}&=&\lan \sigma_2(Q)QR\ran=
   \frac{1}{2}(v^1(v^2)^2)_X-\frac{1}{6}(v^2(v^1)^3)_X+2(v^2v^3(\log v^3-1))_X+
   v^1_Xv^1v^3\log v^3\\
   &&+(v^1)^2v^3_X\log v^3+\frac{1}{2}(\frac{1}{2}(v^1)^2v^2+(v^2)^2\log v^3)_X,\\
 \frac{1}{2}\frac{\pa v^3}{\pa T^{Q,2}}&=&\lan
 \sigma_2(Q)PR\ran=\left[\frac{1}{6}(v^2)^3+\frac{1}{12}(v^1)^3v^3+v^1v^2v^3\log v^3+
 (v^3)^2(\log v^3-2)\right]_X.
 \eean
From above, we can extract the following two-point functions:
 \bean
 \lan \sigma_1(Q)P\ran&=&\frac{1}{2}v^1v^2-\frac{1}{12}(v^1)^3+v^3(\log v^3-1),\\
 \lan \sigma_1(Q)Q\ran&=&\frac{1}{4}(v^1)^2+v^2\log v^3,\\
 \lan \sigma_1(Q)R\ran&=&\frac{1}{4}(v^2)^2+\frac{1}{2}v^1v^3\log v^3,\\
 \lan \sigma_2(Q)P\ran&=&\frac{1}{60}(v^1)^5-\frac{1}{6}(v^1)^3v^2+\frac{1}{2}(v^1)^2v^3+
 \frac{1}{2}v^1(v^2)^2+2v^2v^3(\log v^3-1),\\
 \lan \sigma_2(Q)Q\ran&=&-\frac{1}{24}(v^1)^4+\frac{1}{2}v^2(v^1)^2+(v^2)^2\log v^3+2v^1v^3(\log v^3-1),\\
 \lan \sigma_2(Q)R\ran&=&\frac{1}{6}(v^2)^3+\frac{1}{12}(v^1)^3v^3+v^1v^2v^3\log v^3+
 (v^3)^2(\log v^3-2).
  \eean
\newpage
\subsection{Primary $T^{R,n}$-flows}
 \bean
 \frac{1}{2}\frac{\pa v^1}{\pa T^{R,0}}&=&\frac{1}{2}v^2_X,\\
 \frac{1}{2}\frac{\pa v^2}{\pa T^{R,0}}&=&(\frac{1}{4}v^1v^2_X+\frac{1}{2}v^3_X),\\
 \frac{1}{2}\frac{\pa v^3}{\pa T^{R,0}}&=&\frac{1}{4}(v^1v^3)_X,\\
 \frac{1}{2}\frac{\pa v^1}{\pa T^{R,1}}&=&\lan \sigma_1(R)PQ\ran=
 \left[\frac{1}{2}v^1v^3+\frac{1}{4}(v^2)^2\right]_X,\\
 \frac{1}{2}\frac{\pa v^2}{\pa T^{R,1}}&=&\lan \sigma_1(R)QR\ran=
 \frac{1}{2}(v^3v^2)_X+\frac{1}{4}(v^1v^1_Xv^3+v^1v^2_Xv^2+(v^1)^2v^3_X),\\
 \frac{1}{2}\frac{\pa v^3}{\pa T^{R,1}}&=&\lan \sigma_1(R)PR\ran=
 \left[\frac{1}{4}(v^3)^2+\frac{1}{4}v^1v^2v^3\right]_X,\\
  \frac{1}{2}\frac{\pa v^1}{\pa T^{R,2}}&=&\lan \sigma_2(R)PQ\ran=
 \left[\frac{1}{6}(v^2)^3+v^2v^1v^3+\frac{1}{2}(v^3)^2\right]_X,\\
   \frac{1}{2}\frac{\pa v^2}{\pa T^{R,2}}&=&\lan \sigma_2(R)QR\ran=
   \frac{1}{2}((v^2)^2v^3)_X+\frac{1}{2}(v^1)^2(v^2v^3)_X+\frac{1}{2}v^1v^1_Xv^2v^3+
   \frac{3}{2}v^1v^3v^3_X\\
   &&+\frac{1}{2}v^1_X(v^3)^2+\frac{1}{4}v^1(v^2)^2v^2_X,\\
 \frac{1}{2}\frac{\pa v^3}{\pa T^{R,2}}&=&\lan
 \sigma_2(R)PR\ran=\left[\frac{1}{4}(v^1)^2(v^3)^2+\frac{1}{4}v^1(v^2)^2v^3+
 \frac{1}{2}v^2(v^3)^2\right]_X.
 \eean
From above, we can extract the following two-point functions:
 \bean
 \lan \sigma_1(R)P\ran&=&\frac{1}{2}v^2v^3,\\
 \lan \sigma_1(R)Q\ran&=&\frac{1}{2}v^1v^3+\frac{1}{4}(v^2)^2,\\
 \lan \sigma_1(R)R\ran&=&\frac{1}{4}(v^3)^2+\frac{1}{4}v^1v^2v^3,\\
 \lan \sigma_2(R)P\ran&=&\frac{1}{2}(v^2)^2v^3+\frac{1}{2}v^1(v^3)^2,\\
 \lan \sigma_2(R)Q\ran&=&\frac{1}{6}(v^2)^3+v^2v^1v^3+\frac{1}{2}(v^3)^2,\\
 \lan \sigma_2(R)R\ran&=&\frac{1}{4}(v^1)^2(v^3)^2+\frac{1}{4}v^1(v^2)^2v^3+
 \frac{1}{2}v^2(v^3)^2.
  \eean

 \section{genus-one corrections for the Hamiltonians}

 \bean
 H_{P,0}&=&\int \frac{t^1}{2}+O(\ep^4),\\
 H_{P,1}&=&\int \left[\frac{1}{2}t^2t^3+\frac{1}{4}(t^1)^2\right]-
 \ep^2\int\left[\frac{t^1_Xt^2_X}{6t^2}+\frac{t^2_Xt^3t^3_X}{12t^2}\right]+O(\ep^4),\\
 H_{P,2}&=&\frac{1}{6}\int \left[\frac{1}{2}(t^1)^3+\frac{1}{2}t^2(t^3)^3+
 3t^1t^2t^3+3(t^2)^2\right]-\ep^2\int\left[
 \frac{5}{24}t^1_Xt^1_X+\frac{t^3t^2_Xt^2_X}{4t^2}+\frac{(t^3)^2t^3_Xt^3_X}{12}\right.\\
 &&\left.+\frac{((t^3)^2+2t^1)t^1_Xt^2_X}{12t^2}+\frac{t^3t^1_Xt^3_X}{4}+
 \left(\frac{1}{3}+\frac{t^1t^3}{12t^2}+\frac{(t^3)^3}{24t^2}\right)t^2_Xt^3_X \right]+O(\ep^4),\\
 H_{R,0}&=&\int \frac{t^2}{2}+O(\ep^4),\\
 H_{R,1}&=&\frac{1}{2}\int\left[t^1t^2+\frac{t^2(t^3)^2}{4}\right]-
 \frac{\ep^2}{4}\int\left[\frac{t^3(t^3_X)^2}{6}+\frac{t^3t^2_Xt^1_X}{3t^2}+
 \frac{t^2_Xt^3_X(t^3)^2}{6t^2}+\frac{(t^2_X)^2}{2t^2}+\frac{t^3_Xt^1_{X}}{3}\right]+O(\ep^4),\\
 H_{R,2}&=&\frac{1}{4}\int\left[t^3(t^2)^2+t^2(t^1)^2+\frac{1}{2}t^2t^1(t^3)^2+
 \frac{1}{16}t^2(t^3)^4\right]-\frac{\ep^2}{4}\int\left[\frac{t^3}{3}t^1_Xt^1_X+
 \left(\frac{t^1}{2t^2}+\frac{3(t^3)^2}{8t^2}\right)t^2_Xt^2_X \right.\\
 &&+\left(\frac{t^2}{4}+\frac{t^1t^3}{6}+\frac{(t^3)^3}{8}\right)t^3_Xt^3_X+
 \left(\frac{5}{3}+\frac{t^1t^3}{3t^2}+\frac{(t^3)^3}{12t^2}\right)t^1_Xt^2_X+
 \left(\frac{t^1}{3}+\frac{5(t^3)^2}{12}\right)t^1_Xt^3_X\\
 &&\left.+\left(\frac{4t^3}{3}+\frac{t^1(t^3)^2}{6t^2}+\frac{(t^3)^4}{24t^2}\right)
 t^2_Xt^3_X\right]+O(\ep^4),\\
 H_{Q,0} &=&\int \frac{t^3}{2}+O(\ep^4), \\
 H_{Q,1} &=& \int \left [ \frac{1}{24} (t^3)^3 +\frac{1}{2} t^1 t^3+
t^2\left(\log t^2-\frac{5}{2}\right) \right]-\ep^2 \int\left[ \frac{1}{12 t^2} t^1_X
t^1_X + \frac{(t^3)^2(t^3_X)^2}{48 t^2} \right. \\
 &&+ \left.\frac{t^2_Xt^3_X}{12t^2}+\frac{(t^2_X)^2 t^3}{24(t^2)^2}
+\frac{t^3 t^3_Xt^1_{X}}{12 t^2}\right]+O(\ep^4), \\
 H_{Q,2} &=& \int
\left[\frac{(t^3)^5}{320}+\frac{t^1(t^3)^3}{24}+\frac{(t^1)^2t^3}{4}+\frac{1}{4}(t^3)^2t^2
\left(\log t^2-\frac{3}{2}\right)+t^1t^2\left(\log t^2-\frac{5}{2}\right)\right] \\
&&-\ep^2 \int\left[\frac{1}{6t^2}\left(\frac{1}{8}
(t^3)^2+\frac{t^1}{2}\right)t^1_Xt^1_X+\frac{1}{4t^2}\left(\log t^2+\frac{t^1
t^3}{6t^2}+\frac{(t^3)^3}{24t^2}-\frac{3}{2}\right)t^2_Xt^2_X \right.\\
 &&+\left. \frac{t^3}{12}\left(\log t^2+ \frac{(t^3)^3}{16 t^2}+
\frac{t^1t^3}{4t^2}-\frac{5}{4}\right)t^3_Xt^3_X+
 \frac{t^3}{6t^2}\left(\log t^2-\frac{1}{2}\right)t^1_Xt^2_X \right. \\
&&+ \left. \frac{1}{6}\left(\log t^2+ \frac{(t^3)^3}{8t^2}
+\frac{t^1 t^3}{2t^2}-1\right)t^1_Xt^3_X +\frac{1}{12t^2}\left(\frac{-3(t^3)^2}{4}
+t^1 +(t^3)^2 \log t^2\right)t^2_Xt^3_X \right]+O(\ep^4).
\eean


\end{document}